\documentclass[preprint,aps]{revtex4}
\usepackage{amsmath}

\newcommand{\be}{\begin{equation}}
\newcommand{\ee}{\end{equation}}
\newcommand{\ba}{\begin{eqnarray}}
\newcommand{\ea}{\end{eqnarray}}

\begin{document}

\title{An alternative approach to exact wave functions for time-dependent
coupled oscillator model of charged particle in variable magnetic field}
\author{ Salah Menouar$^{1}$, Mustapha Maamache$^{1}$  and Jeong Ryeol Choi$^{2}$ \vspace{0.7cm}}

\begin{abstract}
\indent .
\end{abstract}

\address
{$^{1}$Laboratoire de Physique Quantique et Syst\`{e}mes
Dynamiques,  D\'{e}partement de Physique, Facult\'{e} des Sciences,
 Universit\'{e} Ferhat Abbas de S\'{e}tif, S\'{e}tif
19000, Algeria}

\address{$^{2}$School of Electrical Engineering and Computer Science,
Kyungpook National University, 1370 Sankguk-dong,
Buk-gu, Daegu 702-701, Republic of Korea}

\begin{center}
\bigskip {\Large An alternative approach to exact wave functions for
time-dependent coupled oscillator model of a charge in variable magnetic
field \vspace{0.4cm}}

$\mathbf{Salah\ Menouar}^{1}\mathbf{,Mustapha}$ $\mathbf{Maamache}^{1}%
\mathbf{\ and}$ $\mathbf{Jeong\ Ryeol\ Choi}^{2}$ \vspace{0.1cm}

\ $^{1}$\textit{\ Laboratoire de Physique Quantique et Syst\`{e}mes
Dynamiques, }

\textit{\ \ \ \ \ \ \ D\'{e}partement de Physique, Facult\'{e} des Sciences,
}

\textit{\ \ \ \ \ \ Universit\'{e} Ferhat Abbas de S\'{e}tif, S\'{e}tif
19000, Algeria }

\textit{\ \ (E-mail: menouar\_salah@yahoo.fr\vspace{0.2cm})}

$\ \ ^{2}$\textit{School of Electrical Engineering and Computer Science,}

\ \ \ \ \textit{Kyungpook National University, 1370 Sankguk-dong, }

\ \ \ \ \textit{Buk-gu, Daegu 702-701, Republic of Korea}\ \ \

\ \ $\ $(\textit{E-mail: choiardor@hanmail.net)}
\end{center}

{\Large Abstract \bigskip }

The quantum states of time-dependent coupled oscillator model for charged
particles subjected to variable magnetic field are investigated using the
invariant operator methods. To do this, we have taken advantage of an
alternative method, so-called unitary transformation approach, available in
the framework of quantum mechanics, as well as a generalized canonical
transformation method in the classical regime. The transformed quantum
Hamiltonian is obtained using suitable unitary operators and is represented
in terms of two independent harmonic oscillators which have the same
frequencies as that of the classically transformed one. Starting from the
wave functions in the transformed system, we have derived the full wave
functions in the original system with the help of the unitary operators. One
can easily take a complete description of how the charged particle behaves
under the given Hamiltonian by taking advantage of these analytical wave
functions.\newline


\ \ \ \ \ \ \ \ \ \ \ \ \ \ \ \ \ \ \ \emph{\ \emph{S}chr\"{o}dinger
equation. }


\newpage

\section{Introduction}

Since the pioneering works of Lewis\cite{lewis1,lewis2,lewis3},
the investigation of the quantum behavior for time-dependent harmonic
oscillator has attracted considerable interest in the literature because it
offers exactly soluble models for dynamical systems in different areas of
physics. There are diverse kind of time-dependent harmonic oscillators such
as Caldirola-Kanai oscillator\cite{ck1,ck2}, parametric oscillator\cite{lo},
and harmonic oscillator with a strongly pulsating mass\cite{sha}, The
two-dimensional time-dependent harmonic oscillator also became an hot topic
in both classical and quantum mechanics. The higher dimensional harmonic
oscillator has played an important role in, for examples, nuclear shell
structure and models of quark confinement. According to the progress of
research for these systems, a relevant question has been naturally raised:
what would happens if two-dimensional harmonic oscillator is constructed by
the coupling of external additive potentials? The investigation for this
issue was initiated by Kim et al.\cite{kim1,kim2,kim3,kim4,kim5,kim6} about
thirty years ago. They considered two harmonic oscillators that are
mutually-coupled so that the resulting potential becomes $V(x_{1},x_{2})=%
\frac{1}{2}\left(c_{1}x_{1}^{2}+c_{2}x_{2}^{2} +c_{3}x_{1}x_{2}\right)$.
They determined the corresponding density matrix rigorously in order to
establish the Wigner function and some of other useful functions in physics.
There are plenty of physical systems and models described by coupled
harmonic oscillators, such as the Lee model in quantum field theory \cite{ss}%
, the Bogoliubov transformation model of superconductivity \cite{kim7},
two-mode squeezed light \cite{cm}, the covariant harmonic oscillator model
for theparton picture \cite{kim8}, and some models in molecular physics \cite%
{fi}. One of the main focuses of research fulfilled in connection with
time-dependent coupled oscillators is some specific problems of
time-dependent coupled electric circuit whose closed-form solutions are now
well known\cite{choi4,choi5,choi6,lml2}. And further, the author of Ref.
\cite{benami} have investigated the propagator for a certain class of
time-dependent coupled and driven harmonic oscillators with time-varying
frequencies and masses using path integral methods.

Above all, a charged particle in a strong uniform magnetic field is a
typical example of natural non-commutative system \cite{gd}, which provides
a good starting point when we are going to discuss the quantum Hall effect
\cite{re}. The external magnetic field is indeed an important factor that
affects to the motion of a charged particle in various physical systems.
When a time-dependent magnetic field is exerted on an electron, it is
impossible to reduce the system to be a one-dimensional problem. Instead, it
can be modeled by two-dimensional time-dependent harmonic oscillator due to
the existence of variable magnetic field $B(t)$. Theoretical and
experimental researches have been carried out extensively on the quantum
properties of this system in the past several decades due to its importance
not only in condensed matter physics but also in plasma physics\cite%
{vvd,Ferre,bbascia,abdella1,abnas,choi1,abdella2,hjb,lml1,smen,rkv}.
Especially, the study of charged particle motion driven by external magnetic
field is crucial in investigating magnetic confinement devices for fusion
plasmas (whose subtopics are tokamaks, mirror machines, bumpy tori, and
stellarators) and for space and astrophysical plasmas (whose subtopics are
magnetospheric plasmas of earth and other planets like pulsar)\cite{rkv}.
The wave functions of a free electron with time-dependent effective mass, in
the presence of a variable magnetic field, are derived in both the Landau
and symmetric gauges \cite{choi2}. The propagators of a charged particle
subjected to a time-dependent magnetic field, which propagate the wave
functions in the spacetime, are derived using the linear and the quadratic
invariants \cite{abdella3}.

An interesting problem that would be worth to be dealt with is the charged
particle system that is described by the Hamiltonian which involves the
static coupling $xy$ and dynamic coupling term $p_{x}p_{y}$ under the
presence of magnetic field. This system may exhibit novel features owing to
the existence of the coupling terms and can be regarded as the
generalization of the Hamiltonian model given in Refs. \cite{choi1} and \cite%
{gel1}. Our intention in this paper is to calculate the exact wave function
for time-dependent coupled oscillators in a variable magnetic field within
the framework of the invariant methods. The calculation is based on the use
of the generalized time-dependent canonical transformations and an
alternative time-dependent unitary transformation.

The present paper is organized as follows. Our problem is formulated in
section 2 through a general time-dependent Hamiltonian describing the motion
of a complicate charged particle system. Some remarks necessary in dealing
with our task will also be presented. In section 3, we show how to simplify
the problem associated with the complicate Hamiltonian of our system using
the canonical transformation method. As an alternative approach, unitary
transformation is also applied, in section 4, in order to transform our
complicate Hamiltonian to that of a more simplified harmonic oscillator. The
quantum solution of the system will be investigated in section 5 using the
invariant methods on the basis of the results obtained in section 4. The
concluding remarks are given in the last section.

\section{Formulating the problem}

Let us formulate our problem by introducing a generalized Hamiltonian
describing the motion of a charged particle that have time-dependent
effective mass $m(t)$ in the presence of a variable magnetic field. The
effective mass of charged particles, such as electrons or holes in any
system, may modified through their interaction with surroundings or various
excitations like energy\cite{dvr}, stress\cite{pkl}, temperature\cite{pdj},
and pressure\cite{him}. It is therefore natural to think that the effective
mass varies with time according to the change of the environments. Moreover,
if we vary the external magnetic field randomly in the heterojunctions and
solid solutions, the effective mass of an electron also varies in a random
fashion in response to the fluctuation of the composition in the system\cite%
{zsg}.

We consider the electromagnetic potential in the symmetric gauge such that $%
\overrightarrow{A}\big( \frac{-B(t)}{2}y,\frac{B(t)}{2}x,0\big) $ where $x$
and $y$ are the position operators. Then, for the dynamical system of our
interest, the Hamiltonian has the form%
\begin{equation}
H(x,y,t)=\frac{1}{2m(t)}\left( \pi _{x}^{2}+\pi _{y}^{2}\right) +\frac{1}{2}%
m(t)\varpi ^{2}(t)\left( x^{2}+y^{2}\right) +a(t)xy+b(t)p_{x}p_{y},
\end{equation}%
where $\ \pi _{x}=p_{x}-\frac{eB(t)y}{2}$ and $\pi _{y}=p_{y}+\frac{eB(t)x}{2%
}$ while $p_{x}$ and $p_{y}$ are conjugate momentum operators that are given
by $p_{x}=-i\hbar \partial /\partial x$ and $p_{y}=-i\hbar \partial
/\partial y$.  $\varpi (t)$ is the oscillating frequency that is an
arbitrary function of time. To generalize the problem, we suppose that the
other parameters, $m(t),$ $a(t)$ and $b(t)$, are also arbitrary time
functions.

In terms of $p _{x}$ and $p _{y}$, the Hamiltonian (1) can be rewritten as
\begin{equation}
H(x,y,t)=\frac{1}{2m(t)}\left( p_{x}^{2}+p_{y}^{2}\right) +\frac{1}{2}%
m(t)\omega ^{2}(t)\left( x^{2}+y^{2}\right) +a(t)xy+b(t)p_{x}p_{y}+\frac{%
\varpi _{c}(t)}{2}L_{z}.
\end{equation}%
Here, $\omega (t)$ is a modulation frequency which takes the form $\omega
^{2}(t)=\varpi ^{2}(t)+\frac{\varpi _{c}^{2}(t)}{4}$ where $\varpi _{c}(t)=%
\frac{eB(t)}{M(t)} $ is the Larmor frequency, and $L_{z}=xp_{y}-p_{x}y=-i%
\hbar \left( x\frac{\partial }{\partial y}-y\frac{\partial }{\partial x}%
\right) $ is the canonical angular momentum in the axial $z$ direction. The
study of quantum features for this system may be a fascinating task in its
own right both from a physical and a mathematical point of view. We will
show how to convert this Hamiltonian to a simple form in the following two
sections by means of the canonical transformation and unitary
transformation, respectively. These procedures may enables us to derive the
quantum solutions of the system.

\section{Canonical transformation}

The method of time-dependent canonical transformation is in fact very
powerful in investigating the mechanical behavior of dynamical systems. We
can convert a given Hamiltonian into any desired one through this method. In
order to cast the Hamiltonian of our problem into a more soluble form, we
take the advantage of  the time-dependent canonical transformation $%
(x,y,p_{x},p_{y})\rightarrow $ \ $(X,Y,P_{X},P_{Y}) $ defined as.
\begin{equation}
\binom{x}{y}=\left(
\begin{array}{cc}
\cos \phi (t) & \sin \phi (t) \\
-\sin \phi (t) & \cos \phi (t)%
\end{array}%
\right) \binom{X}{Y}\text{ , }
\end{equation}%
\begin{equation}
\binom{p_{x}}{p_{y}}=\left(
\begin{array}{cc}
\cos \phi (t) & \sin \phi (t) \\
-\sin \phi (t) & \cos \phi (t)%
\end{array}%
\right) \binom{P_{X}}{P_{Y}},
\end{equation}%
where%
\begin{equation}
\phi (t)=-\frac{1}{2}\int \varpi _{c}(t)dt.
\end{equation}%
From the fundamentals of classical mechanics, we have \cite{gold}%
\begin{eqnarray}
p_{x} &=&\frac{\partial }{\partial x}F_{1}\left( x.y,P_{X},P_{Y},t\right)
\text{ \ , }X=\frac{\partial }{\partial P_{X}}F_{1}\left(
x.y,P_{X},P_{Y},t\right) \text{\ , } \\
p_{y} &=&\frac{\partial }{\partial y}F_{1}\left( x.y,P_{X},P_{Y},t\right)
\text{ \ , }Y=\frac{\partial }{\partial P_{Y}}F_{1}\left(
x.y,P_{X},P_{Y},t\right) ,
\end{eqnarray}%
and%
\begin{equation}
(P_{X}\dot{X}+P_{Y}\dot{Y}-H(X,Y,t)=p_{x}\dot{x}+p_{y}\dot{y}-H(x,y,t)+\frac{%
\partial F_{1}}{\partial t},
\end{equation}
where $F_1$ is the generating functions responsible for the transformation.
Through these relations, $F_1$ is easily found to be%
\begin{equation}
F_{1}\left( x.y,P_{X},P_{Y},t\right) =\left( P_{X}\cos \phi +P_{Y}\sin \phi
\right) x+\left( -P_{X}\sin \phi +P_{Y}\cos \phi \right) y ,
\end{equation}%
\begin{equation}
\frac{\partial F_{1}}{\partial t}=-\dot{\phi}(t)L_{z}=-\frac{\varpi _{c}(t)}{%
2}L_{z} .
\end{equation}%
In terms of the new conjugate variables $\left( X,Y,P_{X},P_{Y},\right) $,
the Hamiltonian (2) becomes%
\begin{eqnarray}
H(X,Y,t) &=&\frac{1}{2m_{-}(t)}P_{X}^{2}+\frac{1}{2m_{+}(t)}P_{Y}^{2}+\frac{1%
}{2}m_{-}(t)\omega _{-}^{2}(t)X^{2}+\frac{1}{2}m_{+}(t)\omega
_{+}^{2}(t)Y^{2}  \notag \\
&&+a_{1}(t)XY+b_{1}(t)P_{X}P_{Y} ,
\end{eqnarray}%
where%
\begin{equation}
\frac{1}{m_{-}(t)}=\frac{1}{m(t)}-2b(t)\sin \phi \cos \phi \text{,}
\end{equation}%
\begin{equation}
\frac{1}{m_{+}(t)}=\frac{1}{m(t)}+2b(t)\sin \phi \cos \phi \text{,}
\end{equation}%
\begin{equation}
\omega _{-}(t)=\left( \frac{m(t)\omega ^{2}(t)-2a(t)\sin \phi \cos \phi }{%
m_{-}(t)}\right) ^{1/2}\text{,}
\end{equation}%
\begin{equation}
\omega _{+}(t)=\left( \frac{m(t)\omega ^{2}(t)+2a(t)\sin \phi \cos \phi }{%
m_{+}(t)}\right) ^{1/2}\text{,}
\end{equation}%
\begin{equation}
a_{1}(t)=a(t)\left( \cos ^{2}\phi -\sin ^{2}\phi \right) \text{,}
\end{equation}%
\begin{equation}
b_{1}(t)=b(t)\left( \cos ^{2}\phi -\sin ^{2}\phi \right) .
\end{equation}%
To eliminate the dynamical term $P_{X}P_{Y}$, we take the second canonical
transformation by recasting the canonical variables $(X,Y,P_{X},P_{Y})$ in
terms of new variables ($q_1$, $q_2$, $p_1$, $p_2$):
\begin{equation}
X=\left( \frac{m_{+}(t)}{m_{-}(t)}\right) ^{1/4}\left( \frac{q_{1}+q_{2}}{%
\sqrt{2}}\right) \text{ \ , \ }Y=\left( \frac{m_{-}(t)}{m_{+}(t)}\right)
^{1/4}\left( \frac{-q_{1}+q_{2}}{\sqrt{2}}\right) \text{\ ,}
\end{equation}%
\begin{equation}
P_{X}=\left( \frac{m_{-}(t)}{m_{+}(t)}\right) ^{1/4}\left( \frac{p_{1}+p_{2}%
}{\sqrt{2}}\right) \text{ \ , \ }P_{Y}=\left( \frac{m_{+}(t)}{m_{-}(t)}%
\right) ^{1/4}\left( \frac{-p_{1}+p_{2}}{\sqrt{2}}\right) .
\end{equation}%
The canonical transformation based on Eqs. (18) and (19) enables us to
transform $H(X,Y,t)$ into $H(q_{1},q_{2},t)$. Thus, straightforwardly, we
have
\begin{equation}
H(q_{1},q_{2},t)=\frac{1}{2m_{1}(t)}p_{1}^{2}+\frac{1}{2m_{2}(t)}p_{2}^{2}+%
\frac{1}{2}m_{1}(t)\omega _{1}^{2}(t)q_{1}^{2}+\frac{1}{2}m_{2}(t)\omega
_{2}^{2}(t)q_{2}^{2}+c(t)q_{1}q_{2},
\end{equation}%
where
\begin{equation}
\frac{1}{m_{1}(t)}=\frac{1}{\sqrt{m_{+}(t)m_{-}(t)}}-b_{1}(t),
\end{equation}%
\begin{equation}
\frac{1}{m_{2}(t)}=\frac{1}{\sqrt{m_{+}(t)m_{-}(t)}}+b_{1}(t),
\end{equation}%
\begin{equation}
\omega _{1}(t)=\left( \frac{\frac{1}{2}\sqrt{m_{+}(t)m_{-}(t)}\left[ \left(
\omega _{-}^{2}(t)+\omega _{+}^{2}(t)\right) \right] -a_{1}(t)}{m_{1}(t)}%
\right) ^{1/2},
\end{equation}%
\begin{equation}
\omega _{2}(t)=\left( \frac{\frac{1}{2}\sqrt{m_{+}(t)m_{-}(t)}\left[ \left(
\omega _{-}^{2}(t)+\omega _{+}^{2}(t)\right) \right] +a_{1}(t)}{m_{2}(t)}%
\right) ^{1/2},
\end{equation}%
\begin{equation}
c(t)=\frac{1}{2}\sqrt{m_{+}(t)m_{-}(t)}\left( \omega _{-}^{2}(t)-\omega
_{+}^{2}(t)\right) .
\end{equation}

Now, to remove the static coupling term $q_{1}q_{2},$ we take another
canonical transformation by introducing the variables ($Q_i$, $P_i$) where $%
i=1,2$, such that \cite{choi5,choi6,benami}
\begin{equation}
\binom{q_{1}}{q_{2}}=\left(
\begin{array}{cc}
\frac{1}{\sqrt{m_{1}(t)}}\cos \frac{\theta (t)}{2} & \frac{1}{\sqrt{m_{1}(t)}%
}\sin \frac{\theta (t)}{2} \\
-\frac{1}{\sqrt{m_{2}(t)}}\sin \frac{\theta (t)}{2} & \frac{1}{\sqrt{m_{2}(t)%
}}\cos \frac{\theta (t)}{2}%
\end{array}%
\right) \binom{Q_{1}}{Q_{2}},
\end{equation}%
\begin{equation}
\binom{p_{1}}{p_{2}}=\left(
\begin{array}{cc}
\sqrt{m_{1}(t)}\cos \frac{\theta (t)}{2} & \sqrt{m_{1}(t)}\sin \frac{\theta
(t)}{2} \\
-\sqrt{m_{2}(t)}\sin \frac{\theta (t)}{2} & \sqrt{m_{2}(t)}\cos \frac{\theta
(t)}{2}%
\end{array}%
\right) \binom{P_{1}}{P_{2}}-\left(
\begin{array}{cc}
\frac{\dot{m}_{1}(t)}{2} & 0 \\
0 & \frac{\dot{m}_{2}(t)}{2}%
\end{array}%
\right) \binom{q_{1}}{q_{2}},
\end{equation}
where $\theta(t)$ is an arbitrary phase which shall be appropriately
determined afterwards. Equations (26) and (27) do not always represent
canonical transformation\cite{gold} between the variables $\left(
q_{i},p_{i}\right) $ and variables $\left( Q_{i},P_{i}\right) $. If $\left(
Q_{i},P_{i}\right) $ are canonical coordinates, there should exist a new
Hamiltonian which is determined only by the Hamiltonian given in (20) and
the linear transformation of (26) and (27).

The variables $\left( q_{i},p_{i}\right) $ and $\left( Q_{i},P_{i}\right) $
in the two representations must satisfy the following relation\cite{gold}
\begin{equation}
\sum_{i=1}^{2}P_{i}\dot{Q}_{i}-H_{Q}=\sum_{i=1}^{2}{}p_{i}\dot{q}_{i}-H_{q}+%
\frac{\partial F}{\partial t},
\end{equation}%
provided that the transformation is canonical, where $F$ is called a
generating function, possibly a time-dependent function in phase space. From
the equations known in classical mechanics
\begin{equation}
p_{i}=\frac{\partial }{\partial q_{i}}F\left(
q_{1},q_{2},P_{1},P_{2},t\right) \text{ \ , }Q_{i}=\frac{\partial }{\partial
P_{i}}F\left( q_{1},q_{2},P_{1},P_{2},t\right) ~~~~~~~i=1,2,
\end{equation}
the generating function responsible for the transformation is found to be%
\begin{eqnarray}
F\left( q_{1},q_{2},P_{1},P_{2},t\right) &=&\sqrt{m_{1}(t)}\left( P_{1}\cos
\frac{\theta (t)}{2}+P_{2}\sin \frac{\theta (t)}{2}\right) q_{1}  \notag \\
&&+\sqrt{m_{2}(t)}\left( -P_{1}\sin \frac{\theta (t)}{2}+P_{2}\cos \frac{%
\theta (t)}{2}\right) q_{2}  \notag \\
&&+\frac{1}{2}\left( -\frac{1}{2}\dot{m}_{1}(t)q_{1}^{2}-\frac{1}{2}\dot{m}%
_{2}(t)q_{2}^{2}\right) .
\end{eqnarray}

In terms of the new conjugate variables $\left( Q_{i},P_{i}\right) $ the
Hamiltonian of the system can be rewritten as%
\begin{eqnarray}
H_{Q}(Q_{1},Q_{2},t) &=&\frac{1}{2}\left( P_{1}^{2}+P_{1}^{2}\right) +\frac{1%
}{2}\Omega _{1}^{2}(t)Q_{1}^{2}+\frac{1}{2}\Omega _{2}^{2}(t)Q_{2}^{2}
\notag \\
&&+\frac{\dot{\theta}(t)}{2}\left[ P_{1}Q_{2}-P_{2}Q_{1}\right] +\delta
(t)Q_{1}Q_{2},
\end{eqnarray}%
where the time-dependent coefficients $\Omega _{1}(t),\Omega _{2}(t)$ and $%
\delta (t)$ are given by
\begin{equation}
\Omega _{1}(t)=\left( \tilde{\omega}_{1}^{2}(t)\cos ^{2}\frac{\theta (t)}{2}+%
\tilde{\omega}_{2}^{2}(t)\sin ^{2}\frac{\theta (t)}{2}-\frac{c(t)\sin \theta
(t)}{\sqrt{m_{1}(t)m_{2}(t)}}\right) ^{1/2}\text{ },
\end{equation}%
\begin{equation}
\Omega _{2}(t)=\left( \tilde{\omega}_{1}^{2}(t)\sin ^{2}\frac{\theta (t)}{2}+%
\tilde{\omega}_{2}^{2}(t)\cos ^{2}\frac{\theta (t)}{2}+\frac{c(t)\sin \theta
(t)}{\sqrt{m_{1}(t)m_{2}(t)}}\right) ^{1/2}\text{ \ },
\end{equation}%
\begin{equation}
\delta (t)=\frac{1}{2}\left( \tilde{\omega}_{1}^{2}(t)-\tilde{\omega}%
_{2}^{2}(t)\right) \sin \theta (t)+\frac{c(t)\cos \theta (t)}{\sqrt{%
m_{1}(t)m_{2}(t)}},
\end{equation}%
with%
\begin{eqnarray}
\tilde{\omega}_{1}^{2}(t) &=&\omega _{1}^{2}(t)+\frac{1}{4}\left( \frac{\dot{%
m}_{1}^{2}(t)}{m_{1}^{2}(t)}-2\frac{\ddot{m}_{1}(t)}{m_{1}(t)}\right) , \\
\tilde{\omega}_{2}^{2}(t) &=&\omega _{2}^{2}(t)+\frac{1}{4}\left( \frac{\dot{%
m}_{2}^{2}(t)}{m_{2}^{2}(t)}-2\frac{\ddot{m}_{2}(t)}{m_{2}(t)}\right) .
\end{eqnarray}
If we choose $\theta (t)=\mathrm{Const.},$ the term involving $P_{1}Q_{2}$
and $P_{2}Q_{1}$ in Eq. (31) disappears and, consequently, we have
\begin{equation}
H_{Q}(Q_{1},Q_{2},t)=\frac{1}{2}\left( P_{1}^{2}+P_{1}^{2}\right) +\frac{1}{2%
}\Omega _{1}^{2}(t)Q_{1}^{2}+\frac{1}{2}\Omega _{2}^{2}(t)Q_{2}^{2}+\delta
(t)Q_{1}Q_{2}.
\end{equation}

It is notable that, with the above canonical transformation, the coupling $%
\delta (t)$ is a function of the parameters of the original system.
Evidently, the separation of variables in equation (37) requires that $%
\delta (t)=0$ , i.e.%
\begin{equation}
c(t)=\frac{1}{2}\left( \tilde{\omega}_{2}^{2}(t)-\tilde{\omega}%
_{1}^{2}(t)\right) \sqrt{m_{1}(t)m_{2}(t)}\tan \theta ,
\end{equation}%
and consequently%
\begin{equation}
\tan \theta =\frac{\sqrt{m_{+}(t)m_{-}(t)}\left( \omega _{-}^{2}(t)-\omega
_{+}^{2}(t)\right) }{\sqrt{m_{1}(t)m_{2}(t)}\left( \tilde{\omega}_{2}^{2}(t)-%
\tilde{\omega}_{1}^{2}(t)\right) }.
\end{equation}
Under this condition, the Hamiltonian of Eq. (37) reduces to
\begin{equation}
H_{Q}(Q_{1},Q_{2},t)=\frac{1}{2}\left( P_{1}^{2}+P_{1}^{2}\right) +\frac{1}{2%
}\Omega _{1}^{2}(t)Q_{1}^{2}+\frac{1}{2}\Omega _{2}^{2}(t)Q_{2}^{2}.
\end{equation}%
This is the sum of two individual Hamiltonians corresponding to the harmonic
oscillators having the time-dependent frequencies $\ \Omega _{1}(t)$ and $%
\Omega _{2}(t)$, respectively, and having masses that are equal to unity.

\section{Unitary transformations}

The unitary transformations in quantum mechanics is analogous to the
canonical transformations in classical mechanics. In this section, the
relationship between the two transformations will be demonstrated and we
confirm how to obtain the quantum-mechanical Hamiltonian from the classical
one. With the consideration of quantum physics, we replace the canonical
variables $\left( x,y\right) $ by quantum operators $\left( \hat{x},\hat{y}%
\right)$, so that the corresponding Hamiltonian is given by%
\begin{equation}
\hat{H}(\hat{x},\hat{y},t)=\frac{1}{2m(t)}\left( \hat{p}_{x}^{2}+\hat{p}%
_{y}^{2}\right) +\frac{1}{2}m(t)\omega ^{2}(t)\left( \hat{x}^{2}+\hat{y}%
^{2}\right) +a(t)\hat{x}\hat{y}+b(t)\hat{p}_{x}\hat{p}_{y}+\frac{\varpi
_{c}(t)}{2}\hat{L}_{z}.
\end{equation}%
In fact, it is not difficult to show the commutation relations $[ \hat{L}_{z}%
\text{ },\text{ }\hat{x}^{2}+\hat{y}^{2}] =0$ and $[ \hat{L}_{z}\text{ },%
\text{ }\hat{p}_{x}^{2}+\hat{p}_{y}^{2}] =0$. We can also check the
non-commutability of $\hat{L}_{z}$ with some other variables: $[ \hat{L}_{z}%
\text{ },\text{ }\hat{p}_{x}\hat{p}_{y}] \neq 0,$ and $[ \hat{L}_{z}\text{ },%
\text{ }\hat{x}\hat{y}] \neq 0$ and consequently $[ \hat{L}_{z}\text{ },%
\text{ }\hat{H}] \neq 0.$ Considering this fact, we are unable to decompose
the Schr\"{o}dinger equation
\begin{equation}
i\hbar \frac{\partial }{\partial t}\Psi (x,y,t)=\hat{H}(\hat{x},\hat{y}%
,t)\Psi (x,y,t),
\end{equation}%
when we would like to simplify it, because the angular momentum operator $%
\hat{L}_{z}$ and the Hamiltonian $\hat{H}(\hat{x},\hat{y},t)$ does not have
the same eigenstates. To overcome this difficult situation, we transform the
Hamiltonian (41) to a simple form by means of appropriate unitary operators.
In the first step, we perform the unitary transformation%
\begin{equation}
\Psi (x,y,t)=\hat{U}(t)\psi ((x,y,t),
\end{equation}%
where $U(t)$ is a unitary operator of the form
\begin{equation}
\hat{U}(t)=\exp \left( -\frac{i\hat{L}_{z}}{2\hbar }\int \varpi
_{c}(t)dt\right) .
\end{equation}%
Under this transformation, the Schr\"{o}dinger equation of original systems
(42) is mapped into%
\begin{equation}
i\hbar \frac{\partial }{\partial t}\psi (x,y,t)=\hat{H}_{1}(\hat{x},\hat{y}%
,t)\psi (x,y,t),
\end{equation}%
where the new Hamiltonian $\hat{H}_{1}(\hat{x},\hat{y},t)$ has the form%
\begin{equation}
\hat{H}_{1}(\hat{x},\hat{y},t)=\frac{1}{2m_{-}(t)}\hat{p}_{x}^{2}+\frac{1}{%
2m_{+}(t)}\hat{p}_{y}^{2}+\frac{1}{2}m_{-}(t)\omega _{-}^{2}(t)\hat{x}^{2}+%
\frac{1}{2}m_{+}(t)\omega _{+}^{2}(t)\hat{y}^{2}+a_{1}(t)\hat{x}\hat{y}%
+b_{1}(t)\hat{p}_{x}\hat{p}_{y}.
\end{equation}%
Note that the term involving $\hat{L}_{z}$ disappeared in the above
equation, This means that the magnetic field is removed when it is viewed
from an appropriate rotating frame $\phi (t)=-\frac{1}{2}\int \varpi
_{c}(t)dt.$

To simplify the Hamiltonian (46), we use two-step unitary transformation
approach. As a first step, we take the following unitary transformation
\begin{equation}
\psi (x,y,t)=\hat{\Lambda}(t)\varphi (x,y,t),
\end{equation}%
where
\begin{equation}
\hat{\Lambda}(t)=\hat{\Lambda}_{1}(t)\hat{\Lambda}_{2}(t),
\end{equation}%
and $\hat{\Lambda}_{1}(t)$ and $\hat{\Lambda}_{2}(t)$ are given by
\begin{equation}
\hat{\Lambda}_{1}(t)=\exp \left[ \frac{i}{2\hbar }(\hat{p}_{x}\hat{x}+\hat{x}%
\hat{p}_{x})\ln \left( \frac{m_{-}(t)}{m_{+}(t)}\right) ^{1/4}\right] \exp %
\left[ \frac{i}{2\hbar }(\hat{p}_{y}\hat{y}+\hat{y}\hat{p}_{y})\ln \left(
\frac{m_{+}(t)}{m_{-}(t)}\right) ^{1/4}\right] ,
\end{equation}%
\begin{equation}
\hat{\Lambda}_{2}(t)=\exp \left[ -\frac{i}{\hbar }\frac{\pi }{4}(\hat{x}\hat{%
p}_{y}-\hat{y}\hat{p}_{x})\right] .
\end{equation}%
Then, we can transform the Hamiltonian (46) using the formula
\begin{equation}
\hat{H}_{2}(\hat{x},\hat{y},t)=\hat{\Lambda}^{-1}(t)\hat{H}_{1}(\hat{x},\hat{%
y},t)\hat{\Lambda}(t)-i\hbar \hat{\Lambda}^{-1}(t)\frac{\partial }{\partial t%
}\hat{\Lambda}(t).
\end{equation}%
After some algebra, we get
\begin{equation}
\hat{H}_{2}(\hat{x},\hat{y},t)=\frac{1}{2m_{1}(t)}\hat{p}_{x}^{2}+\frac{1}{%
2m_{2}(t)}\hat{p}_{y}^{2}+\frac{1}{2}m_{1}(t)\omega _{1}^{2}(t)\hat{x}^{2}+%
\frac{1}{2}m_{2}(t)\omega _{2}^{2}(t)\hat{y}^{2}+c(t)\hat{x}\hat{y}.
\end{equation}%
In the next transformation we will eliminate the coupled static terms $\hat{x%
}\hat{y}.$ To do this we consider the unitary transformation%
\begin{equation}
\varphi (x,y,t)=\hat{V}(t)\chi (x,y,t).
\end{equation}%
Here, $\hat{V}(t)$ is a time-dependent unitary operator of the form%
\begin{equation}
\hat{V}(t)=\hat{V}_{1}(t)\hat{V}_{2}(t)\hat{V}_{3}(t) ,
\end{equation}%
where
\begin{equation}
\hat{V}_{1}(t)=\exp \left[ \frac{i}{2\hbar }(\hat{p}_{x}\hat{x}+\hat{x}\hat{p%
}_{x})\ln \sqrt{m_{1}(t)}\right] \exp \left[ \frac{i}{2\hbar }(\hat{p}_{y}%
\hat{y}+\hat{y}\hat{p}_{y})\ln \sqrt{m_{2}(t)}\right] ,
\end{equation}%
\begin{equation}
\hat{V}_{2}(t)=\exp \left[ -\frac{i}{\hbar }\frac{\theta }{2}(\hat{x}\hat{p}%
_{y}-\hat{y}\hat{p}_{x})\right] ,
\end{equation}%
\begin{equation}
\hat{V}_{3}(t)=\exp -\frac{i}{4\hbar }\left( \dot{m}_{1}(t)\hat{x}^{2}+\dot{m%
}_{2}(t)\hat{y}^{2}\right) .
\end{equation}
Substituting equation (54) in equation (53), we can obtain a transformed
Hamiltonian that is merely the coupling of two harmonic oscillators having
frequencies $\Omega _{1}(t)$ and $\Omega _{2}(t)$ and unit masses:
\begin{eqnarray}
\hat{H}_{3}(\hat{x},\hat{y},t) &=&\hat{V}^{-1}(t)\hat{H}_{2}(\hat{x},\hat{y}%
,t)\hat{V}(t)-i\hbar \hat{V}^{-1}(t)\frac{\partial }{\partial t}\hat{V}(t)
\notag \\
&=&\frac{1}{2}\left( \hat{p}_{x}^{2}+\hat{p}_{y}^{2}\right) +\frac{1}{2}%
\Omega _{1}^{2}(t)\hat{x}^{2}+\frac{1}{2}\Omega _{2}^{2}(t)\hat{y}^{2}.
\end{eqnarray}
At this stage, one can easily confirm that the relation given in equation
(40) is correct, since it is consistent with equation (58). From unitary
operators (44), (49), (50), (55) and (56), we can confirm that $\hat{\Lambda}%
_{1}(t)$ and $\hat{V}_{1}(t)$ are the squeeze operators whereas $\hat{U}(t),$
$\hat{\Lambda}_{2}(t)$ and$\ \hat{V}_{2}(t)$ are the rotation operators with
the angles $\phi (t),$ $\frac{\pi }{4}$ and $\ \frac{\theta }{2}$,
respectively.

Though the original Hamiltonian (41) involves the static coupling term $xy$
and the dynamic coupling term $p_{x}p_{y}$, the transformed Hamiltonian (58)
does not have such terms. Hence we can easily handle equation (58). In the
following section, we establish the quantum solution (wave function) in the
transformed system. And then, we will take the advantage of the unitary
transformation starting form this wave function using the same unitary
operators introduced in this section in order to derive the full wave
functions in the original system.

\section{Quantum solutions}

The problem of the harmonic oscillator with time-dependent mass and
frequency can be transformed to that of the harmonic oscillator via the
associated invariant \cite{lewis3}. It is easy to verify from Liouville-Von
Neumann equation
\begin{equation}
\frac{d\hat{I}}{dt}=\frac{\partial \hat{I}}{\partial t}+\frac{1}{i\hbar }[
\hat{I} , \hat{H}_{3}] =0,
\end{equation}%
that the invariant associated with the transformed Hamiltonian of
two-dimensional harmonic oscillator is given by%
\begin{eqnarray}
\hat{I}(\hat{x},\hat{y},t) &=&\hat{I}(\hat{x},t)+\hat{I}(\hat{y},t)  \notag
\\
&=&\frac{1}{2}\bigg[ \left( \frac{\hat{x}}{\rho _{1}}\right) ^{2}+\big( \rho
_{1}\overset{\cdot }{\hat{x}}-\dot{\rho}_{1}\hat{x}\big) ^{2}\bigg] +\frac{1%
}{2}\bigg[ \left( \frac{\hat{y}}{\rho _{2}}\right) ^{2}+\big( \rho _{2}%
\overset{\cdot }{\hat{y}}-\dot{\rho}_{2}\hat{y}\big) ^{2}\bigg] ,
\end{eqnarray}%
where $\rho _{1}(t)$ and $\rho _{2}(t)$ are c-number quantity satisfying the
auxiliary equations%
\begin{eqnarray}
\ddot{\rho}_{1}+\Omega _{1}^{2}(t) \rho _{1} &=&1/\rho _{1}^{3}, \\
\ddot{\rho}_{2}+\Omega _{2}^{2}(t) \rho _{2} &=&1/\rho _{2}^{3}.
\end{eqnarray}%
In order to make the invariant hermitian, $\hat{I}^\dagger=\hat{I}$, we
choose only the real solution of (61) and (62). Then, we can find the
eigenfunctions $\ \xi _{n_{1}n_{2}}(x,y,t)$ of $\ \hat{I}(\hat{x},\hat{y},t)$%
, which are a complete orthonormal set corresponding to the time-independent
eigenvalues $\lambda _{n_{1}n_{2}}$, from the eigenvalue equation
\begin{equation}
\hat{I}(\hat{x},\hat{y},t)\xi _{n_{1}n_{2}}(x,y,t)=\lambda _{n_{1}n_{2}}\xi
_{n_{1}n_{2}}(x,y,t),
\end{equation}%
where%
\begin{equation}
\lambda _{n_{1}n_{2}}=\hbar \left( n_{1}+\frac{1}{2}\right) +\hbar \left(
n_{2}+\frac{1}{2}\right).
\end{equation}%
By evaluating equation (63) with the use of equation (60), we have the
eigenstates in the form
\begin{eqnarray}
\xi _{n_{1}n_{2}}(x,y,t) &=&\left[ \frac{1}{\pi \hbar
n_{1}!n_{2}!2^{n_{1}+n_{2}}\rho _{1}\rho _{2}}\right] ^{1/2}\times
H_{n_{1}}\left( \frac{x}{\hbar ^{1/2}\rho _{1}}\right) H_{n_{2}}\left( \frac{%
y}{\hbar ^{1/2}\rho _{2}}\right)  \notag \\
&&\times \exp \left[ \frac{i}{2\hbar }\left( \frac{\dot{\rho}_{1}}{\rho _{1}}%
+\frac{i}{\rho _{1}^{2}}\right) x^{2}+\frac{i}{2\hbar }\left( \frac{\dot{\rho%
}_{2}}{\rho _{2}}+\frac{i}{\rho _{2}^{2}}\right) y^{2}\right] ,
\end{eqnarray}%
where $H_{n_{1}}$ and $H_{n_{2}}$ are the usual Hermite polynomial of order $%
n_{1}$ and $n_{2}$, respectively.

The solution of the Schr\"{o}dinger equation%
\begin{equation}
i\hbar \frac{\partial }{\partial t}\chi _{n_{1}n_{2}}(x,y,t)=\hat{H}_{3}(%
\hat{x},\hat{y},t)\chi _{n_{1}n_{2}}(x,y,t),
\end{equation}%
can be written in the form%
\begin{equation}
\chi _{n_{1}n_{2}}(x,y,t)=e^{i\alpha _{n_{1}n_{2}}(t)}\xi
_{n_{1}n_{2}}(x,y,t),
\end{equation}%
where the phase function $\alpha _{n_{1}n_{2}}(t)$ satisfy the equation%
\begin{equation}
\frac{\partial }{\partial t}\alpha _{n_{1}n_{2}}(t)=\frac{1}{\hbar }%
\left\langle \xi _{n_{1}n_{2}}(x,y,t)\right\vert \frac{\partial }{\partial t}%
-\hat{H}_{3}(\hat{x}.\hat{y},t)\left\vert \xi
_{n_{1}n_{2}}(x,y,t)\right\rangle .
\end{equation}
According to equations (67) and (68), the solutions $\chi_{n_{1}n_{2}}
(x,y,t)$ of the transformed Schr\"{o}dinger equation (66) are given by
\begin{eqnarray}
\chi _{n_{1}n_{2}}(x,y,t) &=&e^{i\alpha _{n_{1}n_{2}}(t)}\left[ \frac{1}{\pi
\hbar n_{1}!n_{2}!2^{n_{1}+n_{2}}\rho _{1}\rho _{2}}\right] ^{1/2}\times
H_{n_{1}}\left( \frac{x}{\hbar ^{1/2}\rho _{1}}\right) H_{n_{2}}\left( \frac{%
y}{\hbar ^{1/2}\rho _{2}}\right)  \notag \\
&&\times \exp \left[ \frac{i}{2\hbar }\left( \frac{\dot{\rho}_{1}}{\rho _{1}}%
+\frac{i}{\rho _{1}^{2}}\right) x^{2}+\frac{i}{2\hbar }\left( \frac{\dot{\rho%
}_{2}}{\rho _{2}}+\frac{i}{\rho _{2}^{2}}\right) y^{2}\right] ,
\end{eqnarray}%
where the phase functions takes the form
\begin{equation}
\alpha _{n_{1}n_{2}}(t)=-\left( n_{1}+\frac{1}{2}\right) \int_{0}^{t}\frac{%
dt^{\prime }}{\rho _{1}^{2}(t^{\prime })}-\left( n_{2}+\frac{1}{2}\right)
\int_{0}^{t}\frac{dt^{\prime }}{\rho _{2}^{2}(t^{\prime })}.
\end{equation}

The relation between the wave functions, $\Psi _{n_{1}n_{2}}(x,y,t)$ in
original system described by the Hamiltonian (41)\ and the wave functions $%
\chi _{n_{1}n_{2}}(x,y,t)$\ in the transformed system is%
\begin{eqnarray}
\Psi _{n_{1}n_{2}}(x,y,t) &=&\hat{U}(t)\hat{\Lambda}(t)\hat{V}(t)\chi
_{n_{1}n_{2}}(x,y,t)  \notag \\
&=&\hat{U}(t)\hat{\Lambda}_{1}(t)\hat{\Lambda}_{2}(t)\hat{V}_{1}(t)\hat{V}%
_{2}(t)\hat{V}_{3}(t)\chi _{n_{1}n_{2}}(x,y,t).
\end{eqnarray}%
Using equations (69), (70) and (71), we derive the full wave functions in
the form%
\begin{eqnarray}
\Psi _{n_{1}n_{2}}(x,y,t) &=&\left[ \frac{\left( m_{1}(t)m_{2}(t)\right)
^{1/2}}{\pi \hbar n_{1}!n_{2}!2^{n_{1}+n_{2}}\rho _{1}\rho _{2}}\right]
^{1/2}  \notag \\
&&\times H_{n_{1}}\left( \frac{\left( \eta _{1}\cos \phi +\eta _{2}\sin \phi
\right) x+\left( -\eta _{1}\sin \phi +\eta _{2}\cos \phi \right) y}{\hbar
^{1/2}\rho _{1}}\right) \text{ \ }  \notag \\
&&\times H_{n_{2}}\left( \frac{\left( \mu _{1}\cos \phi +\mu _{2}\sin \phi
\right) x+\left( -\mu _{1}\sin \phi +\mu _{2}\cos \phi \right) y}{\hbar
^{1/2}\rho _{2}}\right)  \notag \\
&&\times \exp \frac{i}{2\hbar }\left[ \left( \frac{m_{-}}{m_{+}}\right)
^{1/2}f_{1}\cos ^{2}\phi +\left( \frac{m_{+}}{m_{-}}\right) ^{1/2}f_{2}\sin
^{2}\phi +\frac{f_{3}}{2}\sin 2\phi \right] x^{2}  \notag \\
&&\times \exp \frac{i}{2\hbar }\left[ \left( \frac{m_{-}}{m_{+}}\right)
^{1/2}f_{1}\cos ^{2}\phi +\left( \frac{m_{+}}{m_{-}}\right) ^{1/2}f_{2}\sin
^{2}\phi -\frac{f_{3}}{2}\sin 2\phi \right] y^{2}  \notag \\
&&\times \exp \frac{i}{2\hbar }\left[ \left( \left( \frac{m_{+}}{m_{-}}%
\right) ^{1/2}f_{2}-\left( \frac{m_{-}}{m_{-}}\right) ^{1/2}f_{1}\right)
\sin 2\phi +f_{3}\cos 2\phi \right] xy  \notag \\
&&\times \exp i\left[ -\left( n_{1}+\frac{1}{2}\right) \int_{0}^{t}\frac{%
dt^{\prime }}{\rho _{1}^{2}(t^{\prime })}-\left( n_{2}+\frac{1}{2}\right)
\int_{0}^{t}\frac{dt^{\prime }}{\rho _{2}^{2}(t^{\prime })}\right] ,
\end{eqnarray}%
where the time-dependent coefficients $f_{1}(t),$ $f_{2}(t),$ $f_{3}(t),$ $%
\gamma (t),$ $\beta (t),$ $\eta _{1}(t),$ $\eta _{2}(t),$ $\mu _{1}(t)$, and
$\mu _{2}(t)$ are given as follows%
\begin{eqnarray}
f_{1}(t) &=&\left( \frac{\gamma }{2}m_{1}+\frac{\beta }{2}m_{2}\right) \cos
^{2}\theta /2+\left( \frac{\gamma }{2}m_{2}+\frac{\beta }{2}m_{1}\right)
\sin ^{2}\theta /2  \notag \\
&&+\sqrt{m_{1}m_{2}}\left( \beta -\gamma \right) \sin \theta /2\cos \theta
/2,
\end{eqnarray}%
\begin{eqnarray}
f_{2}(t) &=&\left( \frac{\gamma }{2}m_{1}+\frac{\beta }{2}m_{2}\right) \cos
^{2}\theta /2+\left( \frac{\gamma }{2}m_{2}+\frac{\beta }{2}m_{1}\right)
\sin ^{2}\theta /2  \notag \\
&&-\sqrt{m_{1}m_{2}}\left( \beta -\gamma \right) \sin \theta /2\cos \theta
/2,
\end{eqnarray}%
\begin{equation}
f_{3}(t)=\left( -\gamma m_{1}+\beta m_{2}\right) \cos ^{2}\theta /2+\left(
\gamma m_{2}-\beta m_{1}\right) \sin ^{2}\theta /2,
\end{equation}%
\begin{equation}
\gamma (t)=\left( \frac{\dot{\rho}_{1}}{\rho _{1}}+\frac{i}{\rho _{1}^{2}}-%
\frac{\dot{m}_{1}(t)}{2}\right) ,
\end{equation}%
\begin{equation}
\beta (t)=\left( \frac{\dot{\rho}_{2}}{\rho _{2}}+\frac{i}{\rho _{2}^{2}}-%
\frac{\dot{m}_{2}(t)}{2}\right) ,
\end{equation}%
\begin{equation}
\eta _{1}(t)=\left( \frac{m_{-}}{m_{+}}\right) ^{1/4}\left( \sqrt{m_{1}/2}%
\cos \theta /2-\sqrt{m_{2}/2}\sin \theta /2\right) ,
\end{equation}%
\begin{equation}
\eta _{2}(t)=\left( \frac{m_{+}}{m_{-}}\right) ^{1/4}\left( -\sqrt{m_{1}/2}%
\cos \theta /2-\sqrt{m_{2}/2}\sin \theta /2\right) ,
\end{equation}%
\begin{equation}
\mu _{1}(t)=\left( \frac{m_{-}}{m_{+}}\right) ^{1/4}\left( \sqrt{m_{1}/2}%
\sin \theta /2+\sqrt{m_{2}/2}\cos \theta /2\right) ,
\end{equation}%
\begin{equation}
\mu _{2}(t)=\left( \frac{m_{+}}{m_{-}}\right) ^{1/4}\left( -\sqrt{m_{1}/2}%
\sin \theta /2+\sqrt{m_{2}/2}\cos \theta /2\right) .
\end{equation}
The final solutions given in (72) are somewhat complicate, but they are very
useful when predicting the evolution of the probability distribution of the
system. A considerable physical significance of such analytical solutions is
that their application in physical system is very flexible, even when the
internal and external situations of the system vary from time to time.
Numerical solutions obtained from, for example, the FDTD (finite difference
time domain)\cite{fdtd} method are however inconvenient as inputs to further
analyses, especially when the parameters of the charged particle system vary
with time like in this case. One can easily take a complete description of
how the charged particle behaves under the given Hamiltonian, by means of
this analytical wave function.

\section{Conclusion}

Though the motion of charged particles in magnetic fields is a fascinating
problem in both quantum and classical view, most of the relevant research is
concentrated on static problem that can be described by time-\textit{in}%
dependent Hamiltonian. In this paper, this problem is generalized to a more
complicated case that is described in terms of time-dependent Hamiltonian,
by supposing that the parameters such as the effective mass of the charged
particle vary explicitly with time in the presence of variable magnetic
field. We have presented an alternative treatment after reducing the problem
related to the charged particle motion to that of the quantal time-dependent
coupled oscillators.

We approached the problem in two ways. In one, we used a time-dependent
generalized canonical transformations which enabled us to transform the
initial classical Hamiltonian (2) to a more simplified one associated with
the two harmonic oscillators having time-dependent frequencies $\Omega
_{1}(t)$ and $\Omega _{2}(t)$. We have taken, as an another way, an
alternative approach based on unitary transformation that allowed us to
transform the quantal Hamiltonian (41) to an equally simple one, but in the
framework of quantum mechanics. To facilitate the derivation of quantum
states, we introduced dynamical invariant operator $\hat{I}(\hat{x},\hat{y}%
,t).$ The eigenvalues and eigenstates of the invariant operator are obtained
using the Liouville-Von Neumann equation.

We derived the exact wave functions of the system on the basis of the fact
that they are the same as the eigenstates of the invariant operator, except
for some time-dependent phase factors $\exp \left[ i\alpha _{n_{1}n_{2}(t)}%
\right]$. The wave functions in the transformed system presented in equation
(69) are relatively simple and expressed in terms of Hermite polynomial.
However, as you can see from equation (72) with equations (73)-(81), the
wave functions in the original system are somewhat complicate. These wave
functions are represented in terms of $\rho_1$ and $\rho_2$ that are the
solutions of some classical equation of motion given in Eqs. (61) and (62),
respectively. If we let $x_{c,1}$ and $x_{c,2}$ as the two independent
classical solutions of the $x$-component transformed Hamiltonian presented
in (58), $\rho_1$ can be written as $\rho_1 = [a_1x_{c,1}^2+a_2
x_{c,1}x_{c,2} +x_{c,2}^2]^{1/2}$ (see Ref. \cite{cje} for rigorous
mathematical proof). Of course, $y$-component two independent solutions $%
y_{c,1}$ and $y_{c,2}$ hold the similar relation with $\rho_2$: $\rho_2 =
[b_1y_{c,1}^2 +b_2y_{c,1} y_{c,2}+b_3 y_{c,2}^2]^{1/2}$. This implies that
we can take the complete knowledge for the behavior of the system within the
scope that quantum mechanics admits when we exactly know the classical
solutions of the \textit{transformed system}, since we did not used any
approxination in the development of our theory.

Though we expressed the wave functions in terms of the classical solutions
of \textit{transformed system}, many authors represent the wave functions
for time-dependent Hamiltonian system in terms of the classical solutions of
\textit{original system}. However, in this case, it is unclear that we can
represent the wave functions using the classical solutions of \textit{%
original system} since the classical solutions in the original system can
not be decoupled into each coordinate component due to the existence of
coupled terms $\hat{x}\hat{y}$ and $\hat{p}_{x}\hat{p}_{y}$ in equation
(41). Even if it were possible to manage the problem using those of \textit{%
original system}, the corresponding mathematical procedure to obtain the
exact wave functions would be very difficult and the results would become
much more complicate. The wave functions we obtained here can be used to
evaluate not only the quantum mechanical expectation values of various
observables such as physical momentum and quantum energy but also
probability densities and fluctuations of the canonical variables.

Finally, we did not considered thermal effects in this work. The quantum
behaviors of the charged particle motion with consideration of thermal
effects may be a good topic as a next task. \newline

\textbf{Acknowledgements} \newline
The work of J. R. Choi was supported by National Research Foundation of
Korea Grant funded by the Korean Government (No. 2009-0077951). \newline

\end{document}